# Changes in mobility choices during the first wave of the COVID-19 pandemic: a comparison between Italy and Sweden

[1]Daniele Giubergia, [2]Elisa Bin, [1]Marco Diana

[1] Department of Environment, Land, and Infrastructure Engineering, Politecnico di Torino, 24 Corso Duca degli Abruzzi, 10129 Turin, Italy

[2] Integrated Transport Research Lab, KTH Royal Institute of Technology, Sweden

Corresponding author:

Daniele Giubergia, daniele.giubergia@studenti.polito.it; daniele.giubergia@icloud.com

Politecnico di Torino - DIATI
Corso Duca degli Abruzzi, 24
10129 TORINO – ITALIA

## Abstract
The spread of COVID-19 disease affected people's lives worldwide, particularly their travel behaviours and how they performed daily activities. During the first wave of the pandemic, spring 2020, countries adopted different strategies to contain the spread of the virus. The aim of this paper is to analyse the changes in mobility behaviours, focusing on the sustainability level of modal choices caused by the pandemic in two countries with different containment policies in place: Italy and Sweden. Survey data uncovered which transport means was the most used for three different trip purposes (grocery shopping, non-grocery shopping and commuting) both before and during the first wave of the pandemic. The variation in the sustainability level of modal choices was then observed through descriptive statistics and significance tests. By estimating three multinomial logistic regression models, one for each trip purpose, we tried to identify which factors, beyond the country, affected the variation in the sustainability level of the modal choice with the beginning of the pandemic. Results show a greater reduction in mobility among the Italian sample compared to the Swedish one, especially for public transit, and a major inclination by Swedes in travelling by foot and by bike compared to Italians, also due to the greater possibility of making trips during the first wave of the pandemic. Finally, perceived safety on public transit seems to have no significant effects on the variation in the sustainability level of the modal choice with the beginning of restrictions. Our results can be used as a starting point for a discussion on how the COVID-19 pandemic affected attitudes and preferences towards the different travel alternatives. Also, in this work we highlighted how people reacted in different ways to an unprecedented situation in two Countries with opposite containment strategies in place.

## 1. Introduction

A novel coronavirus named SARS-CoV-2 was discovered in December 2019 in Wuhan, China (Zhu et al., 2019). Quickly, COVID-19 disease spread worldwide, and the World Health Organization (WHO) declared the pandemic status on the 11$^{th}$ of March 2020. Rapidly, COVID-19 affected people's lives worldwide, particularly their travelling behaviours and how they performed their daily activities. During the first wave of the pandemic, several countries adopted different strategies to contain the spread of the virus. For example, in Europe, some countries implemented strict lockdowns, such as Italy and France. In contrast, others adopted less strict policies, such as the Netherlands with the so-called "intelligent lockdown" (de Haas et al., 2020), and Sweden, where restrictions were the most permissive (Gargoum, Gargoum, 2021; Rumpler et al., 2020).

Many works have analysed the impact of COVID-19 and related containment measures on mobility behaviours in different countries worldwide. For example, the impacts on mobility behaviours, activities and shopping in Australia were studied by Beck and Hensher both in the early stages of the pandemic (Beck, Hensher, 2020a) and after easing restrictions (Beck, Hensher, 2020b). Similar studies, with various methods, have been carried out in Spain (Aloi et al., 2020), the Netherlands (de Haas et al., 2020), Italy (Pepe et al., 2020; Scorrano, Danielis, 2021), India (Pawar et al., 2020), Sweden (Rumpler et al., 2020), USA (Shamshiripour et al., 202034), Japan (Arimura et al., 2020), Turkey (Shakibaei et al., 2020), Poland (Przybyłowski et al., 2021), Canada (Dianat et al., 2021), Germany (Eisenmann et al., 2021), Switzerland (Molloy et al., 2021), etc., whereas others jointly considered more than one country (Abdullah et al., 2020; Bert et al., 2020; Shibayama et al., 2021). All studies detected large changes in mobility behaviours, how people performed their daily activities, and the number of trips made, with some differences among countries. Many of these works made use of surveys to investigate stated and revealed preferences about how to perform activities and the modal choice both before and during the pandemic period (Abdullah et al., 2020; Bert et al., 2020; de Haas et al., 2020; Eisenmann et al., 2021; Scorrano, Danielis, 2021; Shakibaei et al., 2020; Shamshiripour et al., 2020; Shibayama et al., 2021). Some also analysed travel diaries to compare the variations in the travel behaviours (Beck, Hensher, 2020a; de Haas et al., 2020). Other authors monitored the changes using indirect measures (Aloi et al., 2020; Arimura et al., 2020), such as GPS tracking panels (Molloy et al., 2021) or urban noise levels (Rumpler et al., 2020). Sabat et al. also compared the policy response among different countries (Sabat et al., 2020), while Gargoum S. and Gargoum A. compared the different approaches taken by various countries and their effectiveness in containing the spread of the virus (Gargoum, Gargoum, 2021).

While some studies focused on the effects of containment policies on the transmission of the virus (Cartenì et al., 2020; Gargoum, Gargoum, 2021; Hadjidemetriou et al., 2020), this research aims to compare the changes in mobility behaviours in two countries with opposite containment strategy against the virus's spread: Italy and Sweden. To do so, data coming from a mobility survey distributed in these two countries during the first wave of the pandemic has been used.

Two preliminary works carried out by Bin et al. (2021), and Andruetto et al. (2021) used data coming from the same survey. These two analyses studied the variations in the number of trips made at the beginning of the pandemic, the trade-off between virtual and physical activities and the likelihood of keeping the new habits after the pandemic, with a deeper focus on shopping activities in the second article. Here we rather focus on the variations in the modal share in Italy and Sweden caused by the beginning of restrictions. Considering such variations rather than the differences between countries should help in identifying the role of different containment policies beyond the contextual differences between the two countries, e.g. in terms of modal split. To this effect, a series of descriptive analyses and regression models have been developed to understand better how the different approaches taken in these two countries affected the sustainability of modal choices. Moreover, regression models were used to highlight which factors, along with the residence's country, affected the most modal choices before and during the pandemic.

The two countries have been selected because of the extremely different approaches taken during the first wave of the pandemic. Indeed, Italy was the very first country in Europe to have experienced the spread of COVID-19 disease and where restrictions were the strictest. On the other hand, Sweden was the country with the smallest limitations in Europe. During the analysed period, which goes from April 20$^{th}$ to May 31$^{st}$, 2020, the containment measures in Italy changed. To better understand which restrictions were enforced during the analysis period in the two countries, the main measures are described in Table 1.

The structure of this paper consists of four sections: in section 2, we will present how we obtained data on which we made our analyses and what methodology we used; in section 3, results obtained from the analysis will be presented; in section 4, we will discuss these results, and finally, in section 5 we will present our conclusions and some ideas for future research.

*Table 1 Restrictive measures in place during the analysis period (April 20th – May 31st, 2020). (Bin et al., 2021)*

|  | From April 20th to May 4th 2020 | From May 4th to May 31st 2020 |
|---|---|---|
| Italy[1] | There was a lockdown all-over the country: <ul><li>it was not allowed to travel outside the municipality of residence (except for work and health reasons),</li><li>closure of all schools and universities, all non-essential factories, all shops (except grocery stores, pharmacies, convenience stores, newsagents, and petrol stations), all markets, all hair salons, beauticians and barbers; banks, post offices, and public offices were open,</li><li>closure of restaurants, pubs, and cafes (only home delivery allowed),</li><li>closure of parks, running and jogging were allowed.</li></ul> | Starting from May 4th the country partially reopened. It was allowed: <ul><li>to travel in the region of residence for work and health reasons or for visiting relatives, to travel outside the region for work, health and urgent matters reasons or for coming back home,</li><li>to go back to work for workers in manufacturing and construction industry, as well as real estate agents and wholesalers,</li><li>to pick up take away food,</li><li>to access parks.</li></ul> Also, from May 17th most of the commercial activities were allowed to open and it was possible to travel freely within each region while respecting social distancing and using facemasks in public places. |
| Sweden[2] | There were the following restrictions and guidelines: <ul><li>all restaurants, bars, cafés, school dining halls and other venues serving food and beverages had to respect social distancing and customers had to be always seated when consuming,</li><li>public gatherings and public events for more than 50 people were not allowed,</li><li>pharmacies were not allowed to dispense more medications than patients needed for a three-month period,</li><li>to visit the national care homes for the elderly was not allowed,</li><li>it was forbidden to leave home if experiencing any flu-like symptoms (e.g., coughing, cold, fever),</li><li>it was recommended to work from home whenever possible,</li><li>it was strongly recommended to keep social distance of two meters between people when possible.</li></ul> Moreover, from April many companies started a voluntary lockdown of their facilities. | |

[1] http://www.salute.gov.it/portale/nuovocoronavirus/archivioNotizieNuovoCoronavirus.jsp
[2] https://www.government.se/articles/2020/04/s-decisions-and-guidelines-in-the-ministry-of-health-and-social-affairs-policy-areas-to-limit-the-spread-of-the-covid-19-virusny-sida/

## 2. Data collection & Methodology

Data from the mobility survey described by Bin et al. (2021)[3] and Andruetto et al. (2021) have been used in this analysis. The survey was distributed in many countries via the web between April 20th and May 31st, 2020. In this work, we considered only the Italian and the Swedish datasets, as those were the ones with more samples. This was not for chance: collecting data from Sweden and Italy has been prioritised as the two countries had developed significantly different containment strategies and comparing them can provide insights into the effects of restriction policies and behavioural change. People were asked about changes in their travelling behaviour with the beginning of the pandemic, their internet usage, their perceived safety in performing daily activities, and the willingness to maintain new habits after the end of the COVID-19 pandemic. Three trip purposes have been surveyed: commuting trips, grocery shopping trips, and non-grocery shopping trips. For each trip's purposes, interviewees were asked to choose the most frequent travel means that they used both before and during the pandemic to perform their physical trips to reach these activities. Specifically, the before COVID-19 question was "How did you use to travel to perform these activities before the coronavirus outbreak? If you use multimodal travelling, select the mode which covers the most distance and most commonly used". Whereas the question for the COVID-19 period was "How do you travel to perform these activities now? If you use multimodal travelling, select the mode which covers the most distance and most commonly used". In this work, we considered four possible modal choices:

- Motorised private means;
- Public transport;
- Non-motorised private means;
- Other means that were not specified.

---

[3] The survey structure may be found at Appendix C (16-18), at https://etrr.springeropen.com/articles/10.1186/s12544-021-00473-7#appendices

The total number of responses is 654, 428 of which are coming from Italy and 226 coming from Sweden. Since the survey aimed to capture early changes during the first wave of the pandemic when the lockdown was very strict in Italy, fast responses were required. For this reason, the survey was distributed via the internet through social media and among the universities' networks. Therefore, both the Swedish and the Italian samples are not representative of the two populations. Indeed, there is an oversampling of people with a high educational level: more than 84% of the Swedish sample and more than 62% of the Italian sample have at least a bachelor's degree. Females are also over-represented in the Italian sample (almost 61%), while males are the majority of the Swedish sample (about 57%). For more precise information about the samples' composition, the reader is referred to Bin et al. (2021) and Andruetto et al. (2021). It should, however, be noted that the focus of the research was on a limited sample that does not have general validity in the Swedish and Italian populations. However, this research aims to provide basic insights on the effect of restriction policies in place in the two countries.

Through this survey, we derived the percentage of respondents who declared to prefer a certain transport means for a specific trip purpose (named "modal split" in the following for the sake of briefness), in the Italian and the Swedish sample, before and during the first wave of the pandemic.

Then, also by considering the life cycle assessment proposed by Spreafico and Russo (2020), we grouped these modal choices based on their environmental sustainability level. In particular, three possible categories have been set out:

- Environmentally unsustainable transport means, including all the motorised, private transport means (i.e. the first of the above listed modal choices).
- Environmentally sustainable transport means, including public transit, train and trips made on foot or by bike (i.e. the second and third of the above listed modal choices).
- Unknown, encompassing those who reported an un-specified transport means (i.e. the fourth of the above listed modal choices) along with those who did not respond.

For the sake of brevity, we will not specify, from now on, the term "environmental" when referring to these categories, keeping in mind that we always refer to this aspect of sustainability rather than on the fact that public transport could be seen as not sustainable during pandemics due to the increased threat to human health, if not properly managed. After having excluded all observations falling into the last category, we performed a McNemar test (McNemar, 1947) on a series of 2x2 contingency tables containing the pre-COVID-19 behaviours on the rows and the during-COVID-19 behaviours on the columns to highlight significant changes in the sustainability level of the modal choices with the beginning of the pandemic. Three different multinomial logistic regression models were then estimated to understand which factors, along with the home country, increased or decreased the effects of the pandemic on the environmental sustainability of modal choices. Dependent categorical variables of all models were initially based on the following alternatives:

- "Remained unsustainable" if the interviewee chose an unsustainable transport means both before and during the pandemic.
- "Remained sustainable" if the interviewee chose a sustainable transport means both before and during the pandemic.
- "From sustainable to unsustainable", e.g., from public transit to a private transport means.
- "From unsustainable to sustainable", e.g., from a private transport means to public transit or walking.
- "Other" if the interviewee reported "Other" or "N/A" before and/or during the pandemic.

All answers belonging to the latter alternative were not considered in the analysis. The responses not considered vary depending on the different trip purposes: 9% for grocery shopping trips, 41% for non-grocery shopping trips and 43% for commuting trips. Additionally, and due to the number of observations belonging to these categories, we decided to take into account only the first three alternatives for commuting and non-grocery shopping trips, thus leading to two trinomial models, meanwhile, all the first four categories were considered in the case of grocery shopping trips. The independent variables that were considered in these models are:

- Home country, Sweden vs Italy.
- Gender, female vs male.
- Absence of children vs presence of at least one child in the household.
- Presence of at least one elderly vs absence of elderly in the household.
- Education level, degree vs non-degree.
- Being a worker vs being retired/houseman/unemployed.
- Being a student vs being retired/houseman/unemployed.
- Before-pandemic trip duration, metric variable.
- Low safety vs high safety perceived travelling by public transport.
- Low safety vs high safety perceived travelling by car.

It is well known that the built environment has a large impact on modal choices. Nevertheless, two previous works (Bin et al., 2021; Andruetto et al., 2021), which are based on the same database, found no significant relationship between the residential area (urban/rural) and the variation in mobility behaviours due to pandemics. Therefore, we decided not to consider this variable in the present analysis.

Both the bivariate and the multivariate analyses were performed using the software IBM SPSS Statistics 26.0 (IBM, 2019).

This study comes with limitations. Indeed, we considered the analysis period as homogeneous, while in Italy there have been different restrictions, and thus different behaviours, starting from May 4$^{th}$. Furthermore, there are some sampling biases due to the sampling process, as mentioned in chapter 2 Therefore, results and conclusions must be treated with caution and be interpreted as indicative by the policymaker.

# 3. Results

## 3.1. Changes in the most frequently used transport means for different purposes

For each of the three considered trip purposes, we analysed the modal split before and during the pandemic in both the Italian and Swedish samples. The modal split in Italy and Sweden, before and during the pandemic, for the three different trip purposes are respectively shown in Figure 1, Figure 2, and Figure 3.

Concerning commuting trips, before the beginning of restrictions, respondents belonging to the Swedish sample had more sustainable travel patterns compared to the Italian one. Indeed, the percentage of Italian respondents who mostly used motorised private transport means was 61%, whereas public transit stood at 23%. Vice versa, in the Swedish sample, the most used transport means was public transit (46%), while motorised private mobility stood at 24%. Moreover, the percentage of travellers cycling or walking for commuting trips before the pandemic was higher in Sweden (27%) than Italy (13%). During the first pandemic wave, those who answered "N/A" spread in both countries, probably because many did not perform this kind of trip. In the Italian sample, many more people stopped commuting, thus decreasing all transport means, especially for the public transport which dropped from 23% to 3%. In Sweden, the drop in public transport use was even stronger (from 46% to 8%) and compensated by the increase of non-motorised and motorised private means of transport.

In the case of grocery shopping trips, interviewed from the Swedish sample used more public transit before the beginning of restrictions compared to the Italian sample. Anyway, this percentage reduced from 12% to 3% during the pandemic. The most used transport means is the motorised private one, with a greater share in Italy compared to Sweden. During the first wave of the pandemic, motorised private transport means reduced from 79% to 67% in Italy and from 53% to 51% in Sweden. The reduction in the use of motorised private and public transport means is compensated by the increment in trips performed by walking or by bikes and those who answered "N/A" since they probably did not perform such trips anymore.

Furthermore, regarding non-grocery shopping trips, a greater share of Swedish preferred public transit compared to the Italian one before the pandemic. However, in both cases, the use of this transport means decreased greatly with the beginning of restriction: from 12% to 1% in Italy and from 35% to 9% in Sweden. In Italy, all transport means saw a decrease in their use during the first wave of the pandemic, whereas the majority answered "N/A", again most probably due to respondents not conducting any non-grocery shopping trips after the start of the pandemic. Those who answered "N/A" increased during the pandemic also in the Swedish sample. However, those who chose to walk, bike, or use a motorised private transport means increased as well to compensate the reduction in public transit.

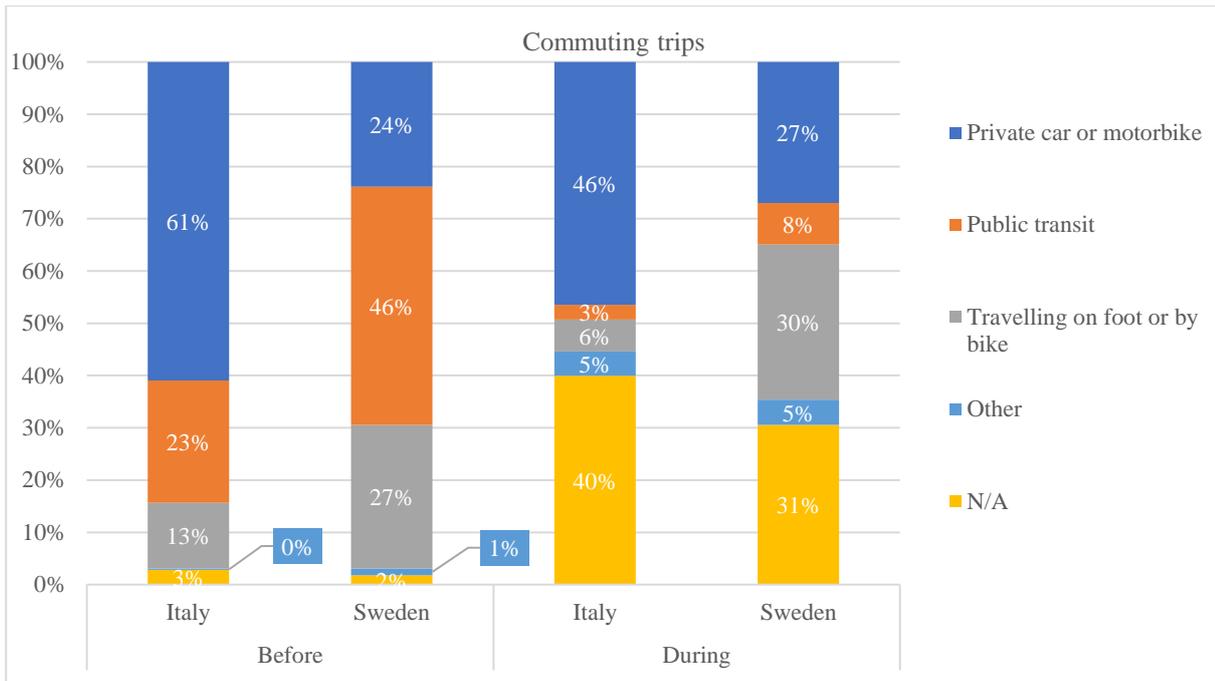

*Figure 1 Modal split for commuting trips*

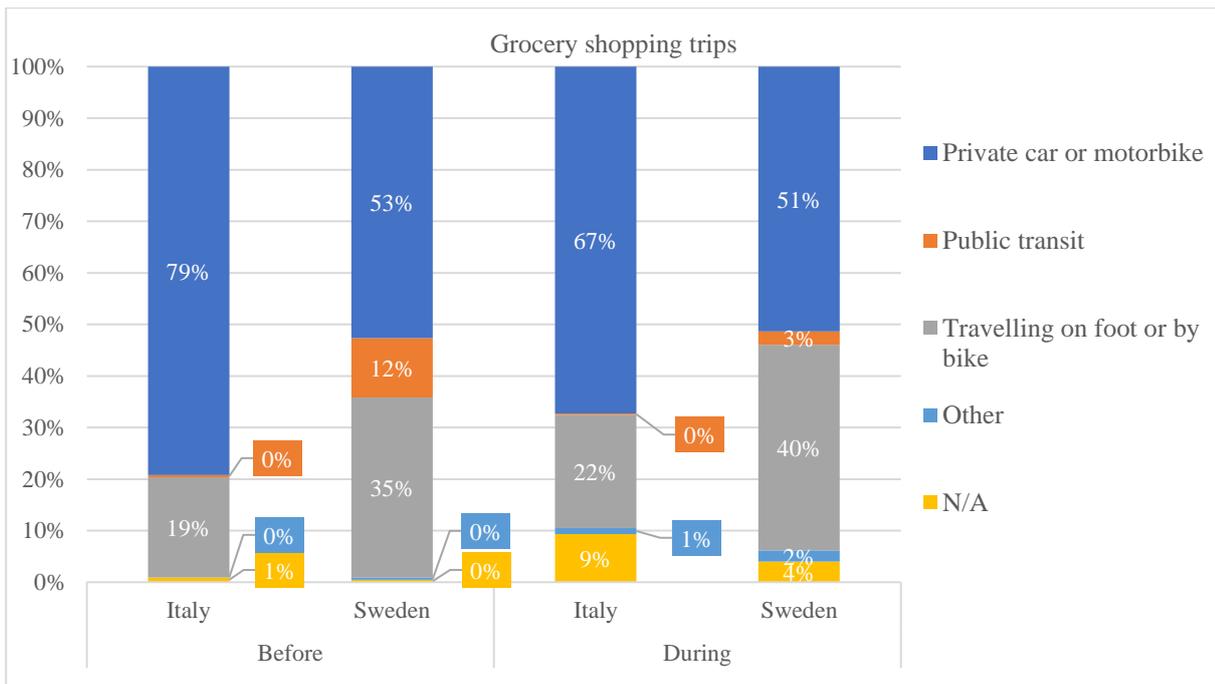

*Figure 2 Modal split for grocery shopping trips*

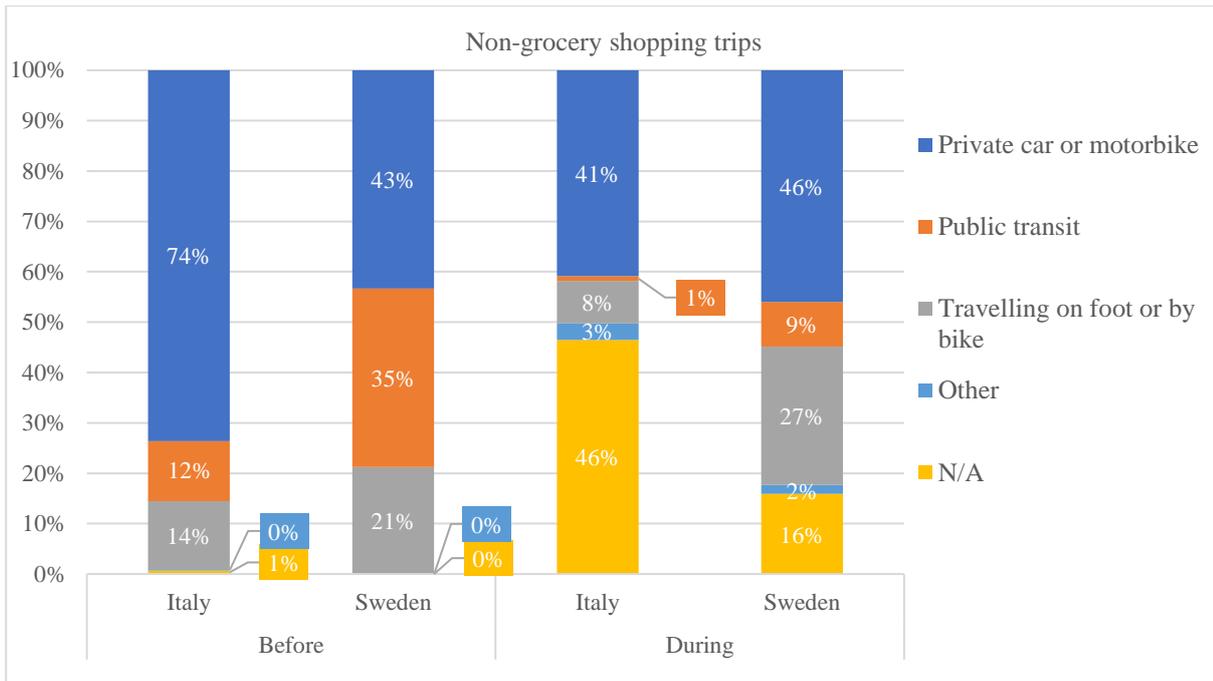

*Figure 3 Modal split for non-grocery shopping trips*

### 3.2. Sustainability patterns of travel choices

After grouping transport means into environmentally sustainable and non-sustainable as explained in the methodology, we derived before/after modal split contingency tables and the related McNemar significance test for different trip purposes (Table 2). The test highlights significant switches between sustainable and unsustainable travel choices with the beginning of the first wave of the pandemic, both in the Italian and in the Swedish sample. The small number of observations falling in some of the categories could be critical in generalising the results. To derive conclusions to the population level must be done with great caution.

In the case of commuting trips, the variation is significant in both countries, with a transition from sustainable to unsustainable mobility. Anyway, it must be highlighted that the percentage of sustainable mobility users is much higher in Sweden than in Italy. Indeed, the percentage of "sustainable commuters" in Italy decreased from 24% to 16%. In the Swedish sample, the reduction has been from 69% to 58%, representing the most frequent category also during the pandemic.

By looking at grocery shopping tables and their corresponding McNemar significance test values, we can see a significant change in the sustainability of the modal choice before and during the pandemic only in the Italian sample. Most of this sample remained unsustainable also after the beginning of restrictions. Interestingly, a sizable part of the sample (12% of total valid answers) shifted from an unsustainable transport means to a sustainable one. Some did the opposite transition, but in the end, those who used sustainable transport means to perform grocery shopping trips increased by 8%.

Concerning non-grocery shopping trips, the variation is significant only in the Swedish sample. In this case, 10% of users switched from sustainable to unsustainable mobility, whereas only 1% did the opposite transition. Therefore, those who used an unsustainable transport means after the beginning of restrictions increased by 9%.

Table 2 Before/After modal split contingency tables and McNemar significance test for different trip purposes.

| Commuting trips in Italy | | Modal choice during pandemic | | Total |
|---|---|---|---|---|
| | | Unsustainable transport means | Sustainable transport means | |
| Modal choice before pandemic | Unsustainable transport means | 174 | 5 | 179 |
| | Sustainable transport means | 25 | 32 | 57 |
| Total | | 199 | 37 | 236 |
| McNemar significance test | | | | p-value=.000 (**) |
| Commuting trips in Sweden | | Modal choice during pandemic | | Total |
| | | Unsustainable transport means | Sustainable transport means | |
| Modal choice before pandemic | Unsustainable transport means | 43 | 2 | 45 |
| | Sustainable transport means | 18 | 81 | 99 |
| Total | | 61 | 83 | 144 |
| McNemar significance test | | | | p-value=.000 (**) |
| Grocery shopping trips in Italy | | Modal choice during pandemic | | Total |
| | | Unsustainable transport means | Sustainable transport means | |
| Modal choice before pandemic | Unsustainable transport means | 271 | 45 | 316 |
| | Sustainable transport means | 17 | 50 | 67 |
| Total | | 288 | 95 | 383 |
| McNemar significance test | | | | p-value=.000 (**) |
| Grocery shopping trips in Sweden | | Modal choice during pandemic | | Total |
| | | Unsustainable transport means | Sustainable transport means | |
| Modal choice before pandemic | Unsustainable transport means | 108 | 7 | 115 |
| | Sustainable transport means | 8 | 89 | 97 |
| Total | | 116 | 96 | 212 |
| McNemar significance test | | | | p-value=1.000 |
| Non-grocery shopping trips in Italy | | Modal choice during pandemic | | Total |
| | | Unsustainable transport means | Sustainable transport means | |
| Modal choice before pandemic | Unsustainable transport means | 164 | 13 | 177 |
| | Sustainable transport means | 11 | 26 | 37 |
| Total | | 175 | 39 | 214 |
| McNemar significance test | | | | p-value=.839 |
| Non-grocery shopping trips in Sweden | | Modal choice during pandemic | | Total |
| | | Unsustainable transport means | Sustainable transport means | |
| Modal choice before pandemic | Unsustainable transport means | 85 | 3 | 88 |
| | Sustainable transport means | 19 | 79 | 98 |
| Total | | 104 | 82 | 186 |
| McNemar significance test | | | | p-value=.001 (**) |

### 3.3. Regression models

Three multinomial logistic regressions, one for each trip purpose, have been estimated. In these three multinomial logistic regression models, we estimated the independent variables' effect on the probability to belong to a certain category among the previously introduced. Those are: "Remained unsustainable", "From sustainable to unsustainable", "From unsustainable to sustainable" (only for grocery trips) and "Remained sustainable", the latter being the reference alternative.

As it can be seen in Table 3, for all the analysed trip purposes living in Sweden rather than in Italy decreases the probability of having used an unsustainable transport means before and during the pandemic, instead of a sustainable transport means (Grocery *Odds Ratio*=0.170, *p-value*<0.05; Non-grocery *Odds Ratio*=0.122, *p-value*<0.05; Commuting *Odds Ratio*=0.088, *p-value*<0.05). Living in Sweden also reduces the probability of having switched from a sustainable transport means to an unsustainable one with the beginning of the pandemic rather than having remained sustainable, for all the analysed trip purposes (Grocery *Odds Ratio*=0.072, *p-value*<0.05; Non-grocery *Odds Ratio*=0.402, *p-value*<0.1; Commuting *Odds Ratio*=0.128, *p-value*<0.05). However, for grocery shopping trips, respondents in Sweden compared to those living in Italy appear to be less likely to have switched from using a car to using a sustainable transport means, rather than using it both before and after the beginning of restrictions (Grocery *O.R.*=0.072; *p-value*<0.05).

Concerning sociodemographic variables, gender seems not to affect the modal choice for those who did not change it during the first wave of the pandemic. However, compared to males, females belonging to the sample were less likely to start using an unsustainable transport means rather than continue to use a sustainable one for commuting trips (Commuting *Odds Ratio*=0.272; *p-value*<0.05).

The absence of children in the household, rather than having at least one, seems to reduce the probability of having used an unsustainable transport means instead of a sustainable one both before and after the pandemic for all the considered trip purposes (Grocery *Odds Ratio*=0.361; *p-value*<0.05; Non-grocery *Odds Ratio*=0.222; *p-value*<0.05; Commuting *Odds Ratio*=0.461; *p-value*<0.05). Furthermore, for all trip purposes, it also appears to decrease the probability of having changed the modal choice from a sustainable to an unsustainable one, rather than having continued to use the public transit or walking or biking after the beginning of restrictions (Grocery *Odds Ratio*=0.404; *p-value*<0.1; Non-grocery *Odds Ratio*=0.185; *p-value*<0.05; Commuting *Odds Ratio*=0.221; *p-value*<0.05).

Compared to those who do not, respondents who have a college degree seem to be less likely to have used a car or a motorcycle instead of sustainable transport means both before and during the pandemic for all trip purposes (Grocery *Odds Ratio*=0.255, *p-value*<0.05; Non-grocery *Odds Ratio*=0.199, *p-value*<0.05; Commuting *Odds Ratio*=0.344, *p-value*<0.05). Having a college degree also looks like it decreases the probability of having changed from a sustainable transport means to an unsustainable one rather than do not change for all trip purposes (Grocery *Odds Ratio*=0.200, *p-value*<0.05; Non-grocery *Odds Ratio*=0.234, *p-value*<0.05; Commuting *Odds Ratio*=0.429, *p-value*<0.1). Finally, it also appears to reduce the probability of having moved from using a car to travel with a sustainable transport means in the case of grocery shopping trips (Grocery *Odds Ratio*=0.298, *p-value*<0.05).

Being a student rather than retired, a houseman/housewife or unemployed looks like reducing the probability of having used unsustainable transport means instead of a sustainable one both before and after the beginning of restrictions for all trip purposes (Grocery *Odds Ratio*=0.118, *p-value*<0.05; Non-grocery *Odds Ratio*=0.095, *p-value*<0.05; Commuting *Odds Ratio*=0.100, *p-value*<0.05). It appears to have no effects on the probability of having shifted from a sustainable modal choice to an unsustainable one for any trip purpose. In the case of grocery shopping trips, it seems to decrease the probability of having done the opposite modal choice shift rather than having continued to travel with sustainable transport means (Grocery *Odds Ratio*=0.283, *p-value*<0.1).

Concerning travel patterns, longer before-pandemic trips seem to enhance the probability of having used an unsustainable transport means rather than a sustainable one both before and after the pandemic for grocery shopping trips (*Odds Ratio*=1.031, *p-value*<0.05). Moreover, both for grocery shopping trips and commuting trips, longer before-pandemic trips appear to increase the probability of having switched from a sustainable to an unsustainable transport means (Grocery *Odds Ratio*=1.036, *p-value*<0.1; Commuting *Odds Ratio*=1.032, *p-value*<0.05).

Respondents who felt not completely safe when travelling by bus have a larger probability of having used an unsustainable transport means instead of a sustainable one both before and during the pandemic for all trip purposes (Grocery *Odds Ratio*=2.912, *p-value*<0.05; Non-grocery *Odds Ratio*=2.302, *p-value*<0.1; Commuting *Odds Ratio*=3.600, *p-value*<0.05). Conversely, it appears to not affect the probability of belonging to the category of those who have changed modal choice.

Finally, perceiving travelling by car as unsafe seems to decrease the probability of having used an unsustainable transport means instead of a sustainable one both before and during the pandemic for grocery shopping trips (Grocery *Odds Ratio*=0.402, *p-value*<0.05). Always for grocery shopping trips, a lower safety feeling in the car appears to decrease the probability of having moved from an unsustainable transport means to a sustainable one rather than having used it also before the pandemic (Grocery *Odds Ratio*=0.177, *p-value*<0.05). This is due to the fact that those who do not feel safe travelling by car probably used a sustainable transport means already before restrictions.

*Table 3 Multinomial logistic regression for the variation in the sustainability of modal choices for different trip purposes*

| | | Variation in the modal choices [a] | | | | | |
|---|---|---|---|---|---|---|---|
| | | Grocery | | Non-grocery | | Commuting | |
| Alternative | Variable | B | Wald | B | Wald | B | Wald |
| Remained unsustainable | (Constant) | 2.946 | 18.538(**) | 4.15 | 22.177(**) | 2.57 | 8.167(**) |
| | Home country, Sweden vs Italy | -1.77 | 44.104(**) | -2.106 | 35.657(**) | -2.428 | 49.323(**) |
| | Before-pandemic trip duration | 0.031 | 6.063(**) | 0.001 | 0.018 | -0.004 | 0.31 |
| | Gender, female vs male | -0.18 | 0.542 | -0.318 | 1.127 | -0.426 | 1.888 |
| | Absence of children vs presence of at least one child in the household | -1.018 | 13.442(**) | -1.505 | 19.051(**) | -0.775 | 5.007(**) |
| | Presence of at least one elderly vs absence of elderly in the household | -0.05 | 0.014 | 0.587 | 1.114 | 0.351 | 0.422 |
| | Education level, degree vs non-degree | -1.368 | 16.929(**) | -1.616 | 14.776(**) | -1.068 | 8.777(**) |
| | Being a worker vs being retired/houseman/unemployed | -0.043 | 0.009 | -0.118 | 0.051 | -0.118 | 0.041 |
| | Being a student vs being retired/houseman/unemployed | -2.137 | 20.747(**) | -2.349 | 15.180(**) | -2.303 | 11.643(**) |
| | Low safety vs high safety perceived travelling by public transport | 1.069 | 8.723(**) | 0.834 | 3.578(*) | 1.281 | 5.705(**) |
| | Low safety vs high safety perceived travelling by car | -0.839 | 3.861(**) | -0.554 | 1.059 | 0.259 | 0.195 |
| From sustainable to unsustainable | (Constant) | -0.907 | 0.381 | 0.439 | 0.099 | 0.642 | 0.306 |
| | Home country, Sweden vs Italy | -1.47 | 8.381(**) | -0.912 | 2.951(*) | -2.059 | 18.406(**) |
| | Before-pandemic trip duration | 0.036 | 3.629(*) | 0.005 | 0.281 | 0.032 | 12.134(**) |

| Variation in the modal choices [a] | | | | | | | |
|---|---|---|---|---|---|---|---|
| | | Grocery | | Non-grocery | | Commuting | |
| Alternative | Variable | B | Wald | B | Wald | B | Wald |
| | Gender, female vs male | -0.136 | 0.083 | 0.061 | 0.018 | -1.301 | 9.489(**) |
| | Absence of children vs presence of at least one child in the household | -0.905 | 3.242(*) | -1.686 | 10.922(**) | -1.509 | 10.510(**) |
| | Presence of at least one elderly vs absence of elderly in the household | -0.448 | 0.281 | 0.954 | 1.487 | 0.551 | 0.566 |
| | Education level, degree vs non-degree | -1.611 | 10.056(**) | -1.454 | 6.503(**) | -0.847 | 3.236(*) |
| | Being a worker vs being retired/houseman/unemployed | 0.745 | 0.417 | 0.507 | 0.286 | 0.157 | 0.032 |
| | Being a student vs being retired/houseman/unemployed | 0.7 | 0.377 | -0.197 | 0.04 | -0.856 | 0.824 |
| | Low safety vs high safety perceived travelling by public transport | 0.852 | 1.131 | 0.757 | 1.16 | 0.678 | 1.085 |
| | Low safety vs high safety perceived travelling by car | -0.58 | 0.49 | -1.192 | 1.082 | -0.068 | 0.007 |
| From unsustainable to sustainable | (Constant) | 0.435 | 0.183 | - | - | - | - |
| | Home country, Sweden vs Italy | -2.638 | 29.441(**) | - | - | - | - |
| | Before-pandemic trip duration | 0.021 | 1.619 | - | - | - | - |
| | Gender, female vs male | 0.604 | 2.344 | - | - | - | - |
| | Absence of children vs presence of at least one child in the household | -0.64 | 2.349 | - | - | - | - |
| | Presence of at least one elderly vs absence of elderly in the household | 0.257 | 0.225 | - | - | - | - |
| | Education level, degree vs non-degree | -1.211 | 7.603(**) | - | - | - | - |
| | Being a worker vs being retired/houseman/unemployed | 0.646 | 0.971 | - | - | - | - |
| | Being a student vs being retired/houseman/unemployed | -1.261 | 2.966(*) | - | - | - | - |
| | Low safety vs high safety perceived travelling by public transport | 0.348 | 0.371 | - | - | - | - |
| | Low safety vs high safety perceived travelling by car | -1.733 | 4.444(**) | - | - | - | - |
| -2 Log Likelihood | | 746.541 | | 446.494 | | 498.382 | |
| Nagelkerke R square | | 0.347 | | 0.384 | | 0.443 | |
| Number of samples | | 595 | | 384 | | 373 | |
| (a) The reference category is "Remained sustainable" | | | | | | | |

# 4. Discussion on the results

By looking at the results from the descriptive analysis presented at §3.1, a considerable variation in the modal split to perform all kinds of trips can be seen. Indeed, there has been a drastic reduction in public transport use during the first pandemic wave in both the Italian and the Swedish samples. However, a greater share continued to prefer this kind of transport means in the latter sample compared to the former. This phenomenon is due to the observed higher reliance of the Swedish sample, compared to the Italian one, on this modal choice before the pandemic. This last figure is in line with what was stated by Fiorello et al. (2016), who studied the level of use of public transport in different countries. Furthermore, the greater reduction in public transport usage in Italy compared to Sweden seems to be in line with the trends reported by Apple, based on mobile phone devices (Apple Inc, 2021). Indeed, during the first pandemic wave, they estimated about a 90% reduction in public transit use in Italy and about 60% in Sweden. This last figure is also reported by Jenelius and Cebecauer (2020), who monitored the variation in public transport ticket validation in Sweden after the beginning of restrictions.

The descriptive analyses showed how the stricter measures imposed in Italy seem to have produced a bigger reduction in the number of trips made, especially with public transport, rather than in Sweden. Indeed, stricter restrictions and closures emanated in Italy rather than in Sweden are reflected in the higher share of those who stopped travelling, given the higher share of those who did not answer any of the possible modal alternatives. In Italy, the drop in the number of trips made produced a decrease in the use of all travel modes. Only for grocery shopping trips, the percentage of those who walked to the shops increased after the beginning of restriction, probably due to a greater propensity in travelling on foot to reach grocery shops near the house, following the prohibition to travel outside the municipality (see Table 1). Conversely, despite the increase of those who answered "N/A", probably because they did not perform any trip after the beginning of restrictions, all modal choices other than the public transit increased their relative importance or remained constant in Sweden. In particular, the increase of those who preferred to walk or to cycle partially compensated for the lowering of public transit users, remaining however sustainable. Still, a transition, especially for commuting trips, from those who used a sustainable transport means to an unsustainable one in both countries can be observed. However, in this research, we analyse differences that might be due also to different ex-ante conditions, e.g. attitude towards sustainable mobility, besides different policies.

The analysis on the sustainability of the modal choices before and during the pandemic made with contingency tables highlighted a transition, especially for commuting trips, from preferring a sustainable transport means to an unsustainable one in both countries. The percentage of those who did this shift, out of the total of those who travelled both before and during the pandemic, is similar in the Italian and the Swedish sample, namely around 10%-13%. However, impacts are much stronger in Italy, where the reduced use of public transit lowered the already low percentage of those who used sustainable transport before the pandemic. Only 14% of the sample used sustainable transport means both before and during the pandemic to reach their workplace, whereas, in Sweden, the same percentage is 56%. In general, all those studies that analysed the variation in the modal share saw a major reduction in the use of buses and trains compared to private transport means (Abdullah et al., 2020; de Haas et al., 2020; Beck, Hensher, 2020a). For example, in the analysis of the case study of Trieste (Italy), Scorrano and Danielis (2021) also saw a similar transition from public transport users towards private modes. In contrast, Eisenmann et al. (2021) stated that transit was the most affected transport means in Germany, while the private car became more important during the first wave of the pandemic. Molloy et al. (2021) analysed the mode share also two months after easing the restrictions and highlighted how the public transit was still 50% below the reference level, while the car was essentially back to the pre-COVID-19 levels.

Multinomial logistic regression models, presented in §3.3, show how living in Sweden rather than Italy increased the probability to have used sustainable transport means instead of an unsustainable one both before and during the pandemic for all trip purposes. This finding highlights a greater inclination of the Swedish sample to travel using sustainable transport means, maintained during the first wave of the pandemic. Of particular relevance, living in Italy rather than Sweden also increased the probability to have switched to an unsustainable transport means instead of having remained sustainable for all trip purposes. Moreover, for grocery shopping trips, living in Sweden rather than in Italy reduced the probability to have switched from an unsustainable transport means to a sustainable one instead of remaining sustainable. Again, in this analysis, we used the country variable to capture how different policies affected the variation in mobility behaviours. Therefore, a confounding factor is linked to the fact that the two countries are also different from other viewpoints beyond the restriction policies. Finally, Italians who started walking or biking to shops often performed their trip mainly by car before the pandemic, whereas Swedish already used public transit, thus an already sustainable transport means, before the pandemic to a larger extent. An increase in the use of active mobility was also reported by Molloy et al. (2021), who saw cycling as the bigger winner during the first wave of the pandemic, and Scorrano and Danielis al (2021), who also tested future scenarios with increased cycle lanes and pedestrian areas.

Other factors that increased or lowered the pandemic's effects on the modal choice are summarised hereafter. In the case of commuting trips, being a male rather than a female increased the probability of having changed the modal choice from a sustainable transport means to an unsustainable one instead of remaining sustainable. This fact could be attributed to the lower accessibility to private motorised vehicles for women compared to men, as stated by Tiikkaja and Liimatainen (2021), but also to differences in more subjective aspects, such as the attachment to the car, the attitudes towards the public transport, the risk perceptions, etc. A greater propensity in using sustainable transport means from females, both before and during the COVID-19 pandemic, is also reported in other studies such as Shakibaei et al. (2020) and Abdullah et al. (2020).

Before-pandemic trip duration affected the modal choice after the beginning of restrictions for commuting and grocery shopping trips. Indeed, longer trips are associated with a greater probability of having switched from a sustainable transport means, realistically the public transit, to an unsustainable one, probably due to the impossibility to perform these trips by walking or by bike.

The presence of at least one child in the household increased the probability of having used an unsustainable transport means instead of a sustainable one both before and during the pandemic. It also enhanced the likelihood of having switched from the latter to the first category rather than having remained sustainable for all trip purposes. This finding is probably due to the necessity to perform these trips along with children. Hotle et al. (2020) did not report a significant correlation between the presence of children and the modal choice, but only in avoiding public places.

Being a student rather than being retired, being a houseman, or being unemployed negatively affected the probability of having used a private transport means rather than an environmentally sustainable one both before and during the first wave of the pandemic. This result is probably due to the lack of availability of this kind of transport means and the cheapness of public transport fares reserved to the students. Also, the residential location could have played a role since most university students live in university cities/campuses, which usually are well served in transit and bike lanes. Various previous works have already detected a greater propensity of students in using more sustainable travel alternatives, especially non-motorised transport means (Whalen et al., 2013; Zhou et al., 2018). During the pandemic, a similar result was obtained by Hotle et al. (2020), which highlighted a greater propensity by the elderly to avoid public transit.

Many studies identified the perceived safety on transport means as a factor that affected the modal choice during the pandemic (Abdullah et al., 2020; de Haas et al., 2020; Shibayama et al., 2021). Moreover, Hotle et al. (2020) stated that perceiving public transport as unsafe made users avoid using it. Przybyłowski et al. (2021) confirmed that the willingness of passengers to return to public transport largely depends on the safety perceived during the pandemic period. More generally speaking, it is well established in the literature that risk attitudes and perceptions affected mobility behaviours during the COVID-19 pandemic suggesting that regions with risk-averse attitudes are more likely to adjust their behaviour in response to the declaration of a pandemic even before government measures are announced (Chan et al., 2020). Anyway, in this study, the perceived safety did not affect the observed patterns of change of modal choice, but only the modal share before the pandemic. In other words, those who perceived public transit as less safe probably decided not to use them already before the pandemic. In contrast, there is no significant relation with those who changed their modal choice. Indeed, Pawar et al. (2020) state that "although the public transport was rated as the most unsafe among other modes, the mode choice decision did not significantly rely on the safety perception of the commuters during the transition to lockdown phase". Since travelling by bus and walking or biking were jointly considered, we could not study the relationship between perceived safety and the switch from using public transport to walking or cycling.

Besides the different policies in place, other factors could have played a role in the variation of mobility behaviours that are not considered here. One of these could be the trust in the governments and thus the compliance with their recommendations. Indeed, several articles in the literature analysed this phenomenon in the context of the COVID-19 pandemics. Some of them agree that greater trust in the institutions is connected to higher compliance with precautionary measures (Han et al., 2020; Murphy et al., 2020; Shanka et al., 2022). On the other hand, Guglielmi et al. (2020) saw a negative direct effect of confidence in the institution on compliance. Instead, Clark et al. (2020) found the trust in governments relatively unimportant in predicting voluntary compliance.

To sum up, the stricter restriction imposed in Italy seems to have caused greater reductions in the number of trips performed compared to Sweden. Other possible explanations could be attributed to factors linked to the two countries' mobility cultures. One of these could be the different attitudes towards non-motorised transport means, evidenced in our data by the higher share of this alternative in Sweden compared to Italy in the before-COVID-19 situation. Also, the different risk perception on the public transport, the different sense of trust in the government measures and, in general, all those cultural differences which, over the past decades, made Italy a much more auto-oriented country compared to Sweden could have played a role. Unfortunately, most of these confounding factors could not be considered due to the lack of data or the definition of the categories. However, it is clear that in both countries public transit saw the greatest reductions with a transition, especially for commuting trips, to private transport means. The possibility of still performing physical trips in Sweden induced many users to perform their trips using "active mobility" (walking and cycling), therefore remaining sustainable. In Italy, the closure of schools and workplaces, combined with a lower readiness to perform trips walking or by bike, did not allow people to experience "active mobility" in commuting trips such as in Sweden. Nonetheless, in the case of grocery shopping trips, which were the least affected, the percentage of those who preferred walking and biking to the shop increased.

# 5. Conclusions

This study aims to analyse the changes in the modal split that occurred in Italy and Sweden: two countries with opposite strategies to contain the spread of the virus during the first wave of the COVID-19 pandemic. Despite the different approaches, Italy and Sweden had similar outcomes in terms of confirmed cases and deaths[4]. As Gargoum S. and Gargoum A. (2021) point out, Italian restrictive measures have been implemented too late, while in Sweden, the reaction has been prompt but not sufficiently restrictive. This study can help understand how different approaches to contain the spread of an epidemic affect mobility behaviours and, specifically, the variation in the modal split. In particular, by comparing the Italian and the Swedish strategies and by identifying factors that resulted in increasing or lowering the effects of the pandemic on the modal choice, the main outcomes that we observed are:

- a greater reduction in the mobility of the Italian sample compared to the Swedish one, especially for the public transport,
- a major inclination by Swedes in travelling on foot and by bike compared to Italians, also due to the greater possibility in making trips during the first pandemic wave,
- a non-significant effect of the safety perceived on public transit on the variation in the sustainability level of the modal choice with the beginning of restrictions.

The different behaviours observed in the two contexts could also be connected to cultural differences, in addition to the containment policies in place. Especially for the last point, the categories used in this work could have hindered the observation of phenomena highlighted in some of the above-reviewed papers. For example, since we jointly considered public transport and non-motorised private vehicles as sustainable modes, we couldn't highlight transitions between these two alternatives due to different risk perceptions. However, our main goal was to study how people reacted, in terms of mobility behaviours, after an unprecedented situation such as the COVID-19 pandemic, in two countries where the virus containments' approach was opposed. In particular, we think that our results can be seen as a valuable contribution for a discussion on how the COVID-19 modified the attitudes and the perceptions towards the different travel alternatives when different containment approaches are put in place.

Acknowledgements
The authors would like to thank the ITRL for providing us the data used in the analysis and the researchers at ITRL and DAVeMoS (University of Natural Resources and Life Sciences BOKU, Vienna) that designed and distributed the survey.

Research data
The datasets analysed in this study have been provided by ITRL. The data are not publicly available but can be shared by the corresponding author upon request.

Competing interests
The authors have no competing interests to declare.

---

[4] Data source: https://covid19.who.int/

Funding

This research is supported by DIATI (Department of Environment, Land, and Infrastructure Engineering, Politecnico di Torino) and ITRL (Integrated Transport Research Lab, KTH Royal Institute of Technology, Sweden).

Authors contributions

All authors contributed to the ideation and planning of the study.

All authors analysed the data and interpreted the results.

All authors contributed to writing the manuscript and approved its final version.

# References

1. Abdullah, M., Dias, C., Muley, D., Shahin, M., 2020. Exploring the impacts of COVID-19 on travel behavior and mode preferences. *Transportation Research Interdisciplinary Perspectives* 8.
   https://doi.org/10.1016/j.trip.2020.100255
2. Aloi, A., Alonso, B., Benavente, J., Cordera, J., Echániz, E., González, F., Ladisa, C., Lezama-Romanelli, R., Lopez-Parra, Á, Mazzei, V., Perrucci, L., Prieto-Quintana, D., Rodríguez, A., Sañudo, R., 2020. Effects of the COVID-19 lockdown on urban mobility: empirical evidence from the City of Santander (Spain). *Sustainability* 12.
   https://doi.org/10.1016/j.trip.2020.100255
3. Andruetto, C., Bin, E., Susilo, Y., & Pernestål, A., 2021. Transition from physical to online shopping alternatives due to the COVID-19 pandemic. *arXiv preprint arXiv:2104.04061*.
4. Apple Inc, "COVID-19 mobility trends", accessed April 23rd, 2021.
   https://covid19.apple.com/mobility
5. Arimura, M., Ha, T., Okumura, K., Asada, T., 2020. Changes in urban mobility in Sapporo city, Japan due to the Covid-19 emergency declarations. *Transportation Research Interdisciplinary Perspectives* 7.
   https://doi.org/10.1016/j.trip.2020.100212
6. Beck, M. J., Hensher, D. A., 2020a. Insight into the Impact of COVID-19 on Household Travel, Working, Activities and Shopping in Australia – the early days under Restrictions. *Transport Policy* 96.
   https://doi.org/10.1016/j.tranpol.2020.07.001
7. Beck, M. J., Hensher, D. A., 2020b. Insight into the Impact of COVID-19 on Household Travel and Activities in Australia – the early days of Easing Restrictions. *Transport Policy* 99.
   https://doi.org/10.1016/j.tranpol.2020.08.004
8. Bert, J., Schellong, D., Hagenmaier, M., Hornstein, D., Wegscheider A. K., Palme, T., 2020. How COVID-19 will shape urban mobility. *BCG*.
   https://www.bcg.com/it-it/publications/2020/how-COVID-19-will-shape-urban-mobility
9. Bin, E., Andruetto, C., Susilo, Y., Pernestål, A., 2021. The trade-off behaviours between virtual and physical activities during COVID-19 pandemic period. *European Transport Research Review* 13.
   https://doi.org/10.1186/s12544-021-00473-7

10. Cartenì, A., Di Francesco, L., Martino, M., 2020. How mobility habits influenced the spread of the COVID-19 pandemic: Results from the Italian case study. *Science of the Total Environment* 741.
https://doi.org/10.1016/j.scitotenv.2020.140489
11. Chan, H.F., Skali, A., Savage, D.A., Stadelmann, D., Torgler, B., 2020. Risk attitudes and human mobility during the COVID-19 pandemic. Sci. Rep. 10, 1–13.
https://doi.org/10.1038/s41598-020-76763-2
12. Clark, C., Davila, A., Regis, M., Kraus, S., 2020. Predictors of COVID-19 voluntary compliance behaviors: An international investigation. *Glob. Transitions* 2, 76–82.
https://doi.org/10.1016/j.glt.2020.06.003
13. de Haas, M., Faber, R., Hamersma, M., 2020. How COVID-19 and the Dutch 'intelligent lockdown' change activities, work and travel behaviour: Evidence from longitudinal data in the Netherlands. *Transportation Research Interdisciplinary Perspectives* 6.
https://doi.org/10.1016/j.trip.2020.100150
14. Dianat, A., Hawkins, J., Habib, K., N., 2021. Assessing the impacts of COVID-19 on activity-travel scheduling: a survey in the Greater Toronto Area. Preprint.
15. Eisenmann, C., Nobis, C., Kolarova, V., Lenz, B., Winkler, C., 2021. Transport mode use during the COVID-19 lockdown period in Germany: The car became more important, public transport lost ground. *Transport policy* 103.
https://doi.org/10.1016/j.tranpol.2021.01.012
16. Fiorello, D., Martino, A., Zani, L., Christidis, P., Navajas-Cawood, E., 2016. Mobility data across the EU 28 member states: results from an extensive CAWI survey. *Transportation research procedia* 14.
https://doi.org/10.1016/j.trpro.2016.05.181
17. Gargoum, S., Gargoum, A., 2021. Limiting mobility during COVID-19, when and to what level? An international comparative study using change point analysis. *Journal of Transport & Health* 20.
https://doi.org/10.1016/j.jth.2021.101019
18. Guglielmi, S., Dotti Sani, G.M., Molteni, F., Biolcati, F., Chiesi, A.M., Ladini, R., Maraffi, M., Pedrazzani, A., Vezzoni, C., 2020. Public acceptability of containment measures during the COVID-19 pandemic in Italy: how institutional confidence and specific political support matter. *Int. J. Sociol. Soc. Policy* 40, 1069–1085.
https://doi.org/10.1108/IJSSP-07-2020-0342
19. Hadjidemetriou, G., Sasidharan, M., Kouyialis, G., Parlikad, A., 2020. The impact of government measures and human mobility trend on COVID-19 related deaths in the UK. *Transportation Research Interdisciplinary Perspectives* 6.
https://doi.org/10.1016/j.trip.2020.100167
20. Han, Q., Zheng, B., Cristea, M., Agostini, M., Belanger, J., Gutzkow, B., Kreienkamp, J., team, P., Leander, P., 2020. Trust in government and its associations with health behaviour and prosocial behaviour during the COVID-19 pandemic. *PsyArXiv Prepr*.
https://doi.org/10.31234/osf.io/p5gns
21. Hotle, S., Murray-Tuite, P., Singh, K., 2020. Influenza risk perception and travel-related health protection behaviour in the US: Insight for the aftermath of the COVID-19 outbreak. *Transportation Research Interdisciplinary Perspectives* 5.
https://doi.org/10.1016/j.trip.2020.100127

22. IBM Corp. Released 2019. IBM SPSS Statistics for Windows, Version 26.0. Armonk, NY: IBM Corp
23. Jenelius, E., Cebecauer, M., 2020. Impacts of COVID-19 on public transport ridership in Sweden: analysis of ticket validations, sales and passenger counts. *Transport Research Interdisciplinary Perspectives* 8.
https://doi.org/10.1016/j.trip.2020.100242
24. McNemar, Q., 1947. Note on the sampling error of the difference between correlated proportions or percentages. *Psychometrika* 12.
https://doi.org/10.1007/BF02295996
25. Molloy, J., Schatzmann, T., Schoeman, B., Tchervenkov, C., Hintermann, B., Axhausen, K. W., 2021. Observed impacts of the Covid-19 first wave on travel behaviour in Switzerland based on a large GPS panel. *Transport Policy* 104.
https://doi.org/10.1016/j.tranpol.2021.01.009
26. Murphy, K., Williamson, H., Sargeant, E., McCarthy, M., 2020. Why people comply with COVID-19 social distancing restrictions: Self-interest or duty? *Aust. New Zeal. J. Criminol.* 53, 477–496.
https://doi.org/10.1177/0004865820954484
27. Pawar, D., Yadav, A., Akolekar, N., Velaga, N., 2020. Impact of physical distancing due to novel coronavirus (SARS-CoV-2) on daily travel for work during transition to lockdown. *Transportation Research Interdisciplinary Perspectives* 7.
https://doi.org/10.1016/j.trip.2020.100203
28. Pepe, E., Bajardi, P., Gauvin, L., Privitera, F., Lake, B., Cattuto, C., Tizzoni, M., 2020. COVID-19 outbreak response: a first assessment of mobility changes in Italy following national lockdown. *MedRxiv*.
https://doi.org/10.1101/2020.03.22.20039933
29. Przybyłowski, A., Stelmak, S., Suchanek, M., 2021. Mobility behaviour in view of the impact of the COVID-19 pandemic – Public transport users in Gdansk case study. *Sustainability*, v.13, iss.364.
https://doi.org/10.3390/su13010364
30. Rumpler, R., Venkataraman, S., Göransson, P., 2020. An observation of the impact of COVID-19 recommendation measures monitored through urban noise levels in central Stocckholm, Sweden. *Sustainable Cities and Society* 63.
https://doi.org/10.1016/j.scs.2020.102469
31. Sabat, I., Neuman-Böhme, S., Varghese, N., E., Barros, P., P., Brouwer, W., van Exel, J., Schreyögg, J., Stargardt, T., 2020. United but divided: Policy responses and people's perceptions in the EU during the COVID-19 outbreak. *Health Policy* 124.
https://doi.org/10.1016/j.healthpol.2020.06.009
32. Scorrano, M., Danielis, R., 2021. Active mobility in an Italian city: Mode choice determinants and attitudes before and during the Covid-19 emergency. *Research in Transportation Economics*.
https://doi.org/10.1016/j.retrec.2021.101031
33. Shakibaei, S., de Jong, G., C., Alpkökin, P., Rashidi, T., H., 2020. Impact of the COVID-19 pandemic on travel behavior in Istanbul: a panel data analysis. *Sustainable Cities and Society* 65.

https://doi.org/10.1016/j.scs.2020.102619

34. Shamshiripour, A., Rahimi, E., Shabanpour, R., Mohammadian, A., 2020. How is COVID-19 reshaping activity-travel behaviour? Evidence from a comprehensive survey in Chicago. *Transport Research Interdisciplinary Perspectives* 7.
https://doi.org/10.1016/j.trip.2020.100216

35. Shanka, M.S., Menebo, M.M., 2022. When and How Trust in Government Leads to Compliance with COVID-19 Precautionary Measures. J. Bus. Res. 139, 1275–1283.
https://doi.org/10.1016/j.jbusres.2021.10.036

36. Shibayama, T., Sandholzer, F., Laa, B., Brezina, T., 2021. Impact of COVID-19 lockdown on commuting: a multi-country perspective. *European Journal of Transport and Infrastructure Research* 21.
https://doi.org/10.18757/ejtir.2021.21.1.5135

37. Spreafico, C., Russo, D., 2020. Exploiting the Scientific Literature for Performing Life Cycle Assessment about Transportation. *Sustainability* 12.
https://doi.org/10.3390/su12187548

38. Tiikkaja, H., & Liimatainen, H., 2021. Car access and travel behaviour among men and women in car deficient households with children. *Transportation Research Interdisciplinary Perspectives*, *10*.
https://doi.org/10.1016/j.trip.2021.100367

39. Whalen, K. E., Páez, A., & Carrasco, J. A., 2013. Mode choice of university students commuting to schooland the role of active travel. *Journal of Transport Geography*, *31*, 132–142.
https://doi.org/10.1016/j.jtrangeo.2013.06.008

40. Zhou, J., Wang, Y., & Wu, J. (2018). Mode choice of commuter students in a college town: An exploratory study from the United States. *Sustainability (Switzerland)*, *10*(9).
https://doi.org/10.3390/su10093316

41. Zhu, N., Zhang, D., Wang, W., Li, X., Yang, B., Song, J., Zhao, X., Huang, B., Shi, W., Lu, R., Niu, P., Zhan, F., Ma, X., Wang, D., Xu, W., Wu, G., Gao, G., Phil, D., Tan, W., et al., 2020. A novel coronavirus from patients with pneumonia in China, 2019. *New England Journal of Medicine* 382.
https://doi.org/10.1056/NEJMoa2001017